\documentclass{cimento}
\usepackage{graphicx}
\usepackage{xspace}
\newcommand{\pt} {\ensuremath{p_\mathrm{T}}\xspace}
\newcommand{\jpsi} {\ensuremath{\rm J/\psi}\xspace}
\newcommand{\trento} {T$_{\rm R}$ENTo\xspace}
\newcommand{\raa} {$R_{AA}$}

\setcopyright{CERN on behalf of the ALICE Collaboration}
\title{Heavy-ion physics with the ALICE detector}
\author{D.~Mi\'skowiec on behalf of the ALICE Collaboration}
\instlist{\inst{}GSI Helmholtzzentrum f\"ur Schwerionenforschung GmbH, Darmstadt, Germany}

\begin{document}

\maketitle

\begin{abstract}
  A selection of recent results from the heavy-ion experiment ALICE at
  the CERN LHC, chosen to address various stages of the
  nucleus-nucleus reaction.
\end{abstract}

\section{Introduction}
\label{sect:intro}
\begin{figure}[b]
\center{\includegraphics[width=0.82\textwidth]{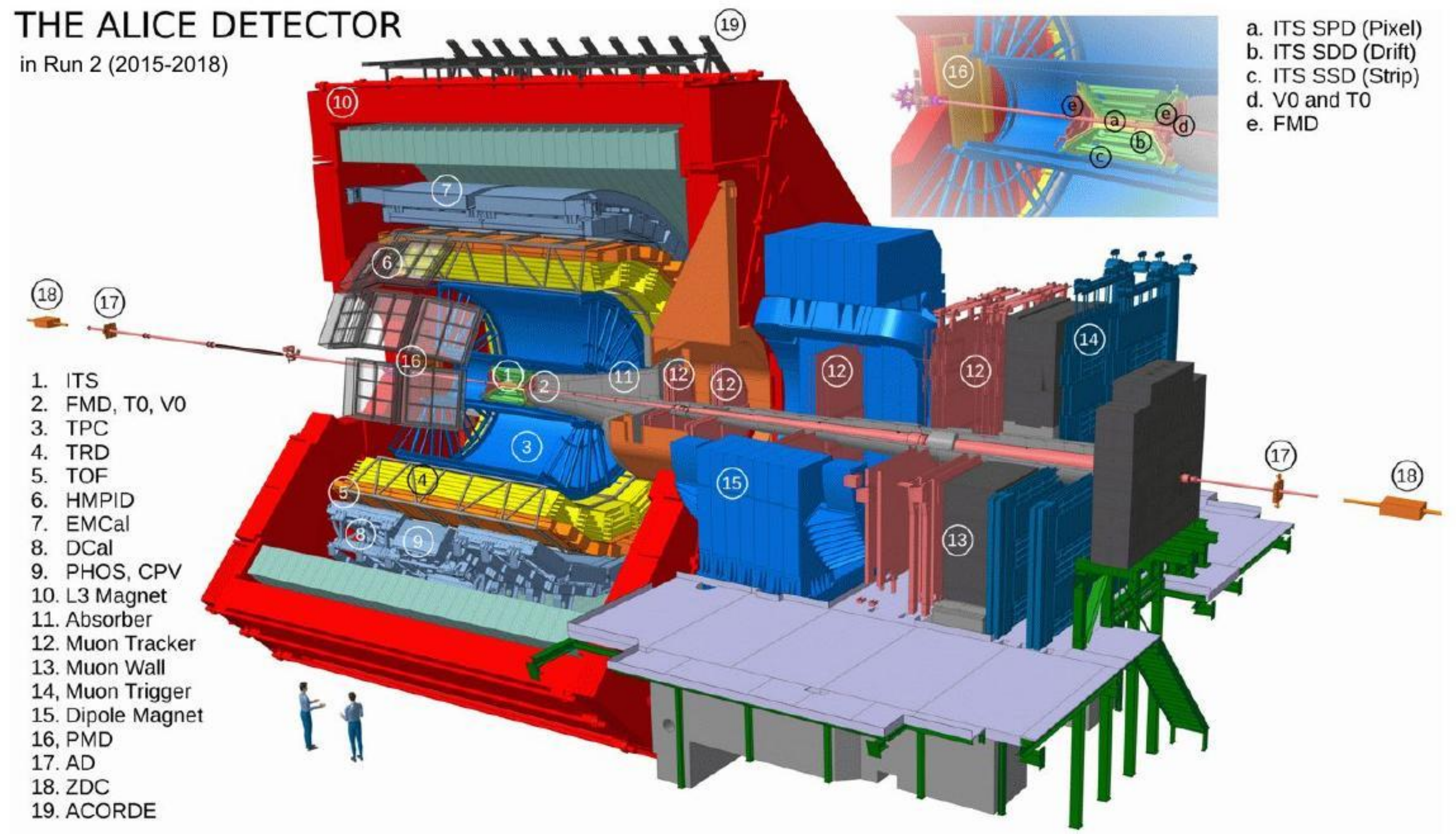}}
\caption{ALICE at the LHC. The central barrel detectors in a solenoid
  magnet cover midrapidity, forward muons are measured after an
  absorber and a dipol magnet, auxilliary detectors provide triggering
  and centrality determination. }
\label{fig:setup}
\end{figure}

It is CERN's mission to ``unite people from all over the world (...)
for the benefit of all.''  The ALICE collaboration at CERN counts 2000
members from 40 countries studying the mechanism of collisions of lead
nuclei at LHC energies and the properties of the special state of matter
created therein. The ALICE apparatus and its performance are described
in Refs. \cite{ref:jinst} and \cite{ref:perf}, respectively.  Its
main components are shown in Fig.~\ref{fig:setup}.
The central barrel detectors provide tracking and particle
identification of charged particles within $-0.9<\eta<0.9$ and
electromagnetic calorimetry within $-0.7<\eta<0.7$. This is
complemented by a muon detector at $-4.0<\eta<-2.5$ as well as several
detectors for triggering and centrality determination.

\begin{figure}[t]
\center{\includegraphics[width=\textwidth]{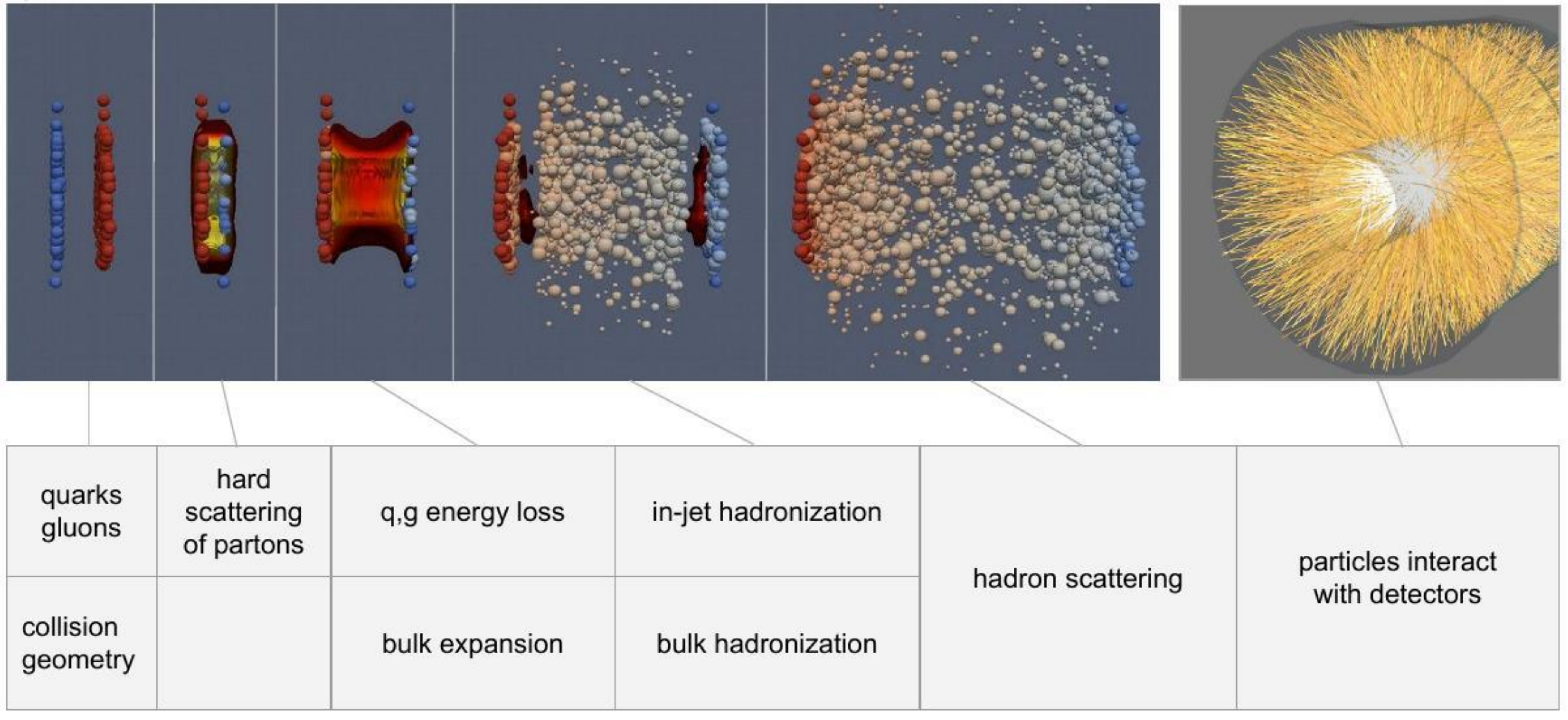}}
\caption{Main stages of a Pb--Pb collision at the LHC. See
  text. Visualization figure taken from Ref.~\cite{ref:madai}. }
\label{fig:scheme}
\end{figure}

The main stages of a Pb--Pb collision at the LHC are schematically
represented in Fig.~\ref{fig:scheme}.
The initial state depends on the collision centrality and on the
distribution of nucleons within the colliding nuclei in the transverse
plane. The nuclei, contracted by a Lorentz factor of $\sim$2000, fly
through each other and the space between their receding disks is
filled with energy in form of gluons and quarks (quark-gluon
plasma). It is helpful, in particular for non-central collision, to
talk about collisions between pairs of nucleons~\cite{ref:czyz}.  The
temporal sequence of these collisions does not depend on their
longitudinal positions within nuclei. Instead, all nucleon-nucleon (NN) 
collisions are initiated at the same time and their duration depends
on their hardness.  Hard collisions finish first and get all the
energy they want; this is why they scale with the number of NN
collisions $N_{\rm coll}$. Soft collisions take longer and compete
among themselves for energy, so they scale like the total available
energy or the number of participating nucleons, $N_{\rm part}$.  Hard
collisions do not compete against soft collisions because they are
faster, and do not compete among themselves because they are rare.

Heavy quarks and high-\pt quarks and gluons are produced in hard
processes.  The ones with a larger than average transverse velocity 
have to traverse a portion of the elongated fireball in the transverse
direction. They scatter and radiate, equilibrating their transverse
velocity with the surrounding medium.  By measuring their energy loss
one learns about the properties of the quark-gluon plasma.

The fireball keeps expanding until the energy density drops below
1~GeV/fm$^3$, at which point the quarks and gluons have to form color
neutral hadrons. This process turns out to depend on the environment.
The formed hadrons interact with each other inelastically, then only
elastically. The chemical and kinetic freeze-out mark the ends of
these two phases and shape the hadron yields and momenta,
respectively. Finally, the hadrons propagate freely until their
interact with our detectors.

In this talk I am briefly discussing several recent (less than 1~year
old) ALICE results, which uniformly address all stages of the nuclear
collisions at the LHC. The reader is referred to the respective ALICE 
publications for details. 

\section{Initial state: Correlation between the elliptic flow and the
  mean transverse momentum}
\label{sect:initial}
\begin{figure}[b]
\center{\includegraphics[width=0.9\textwidth]{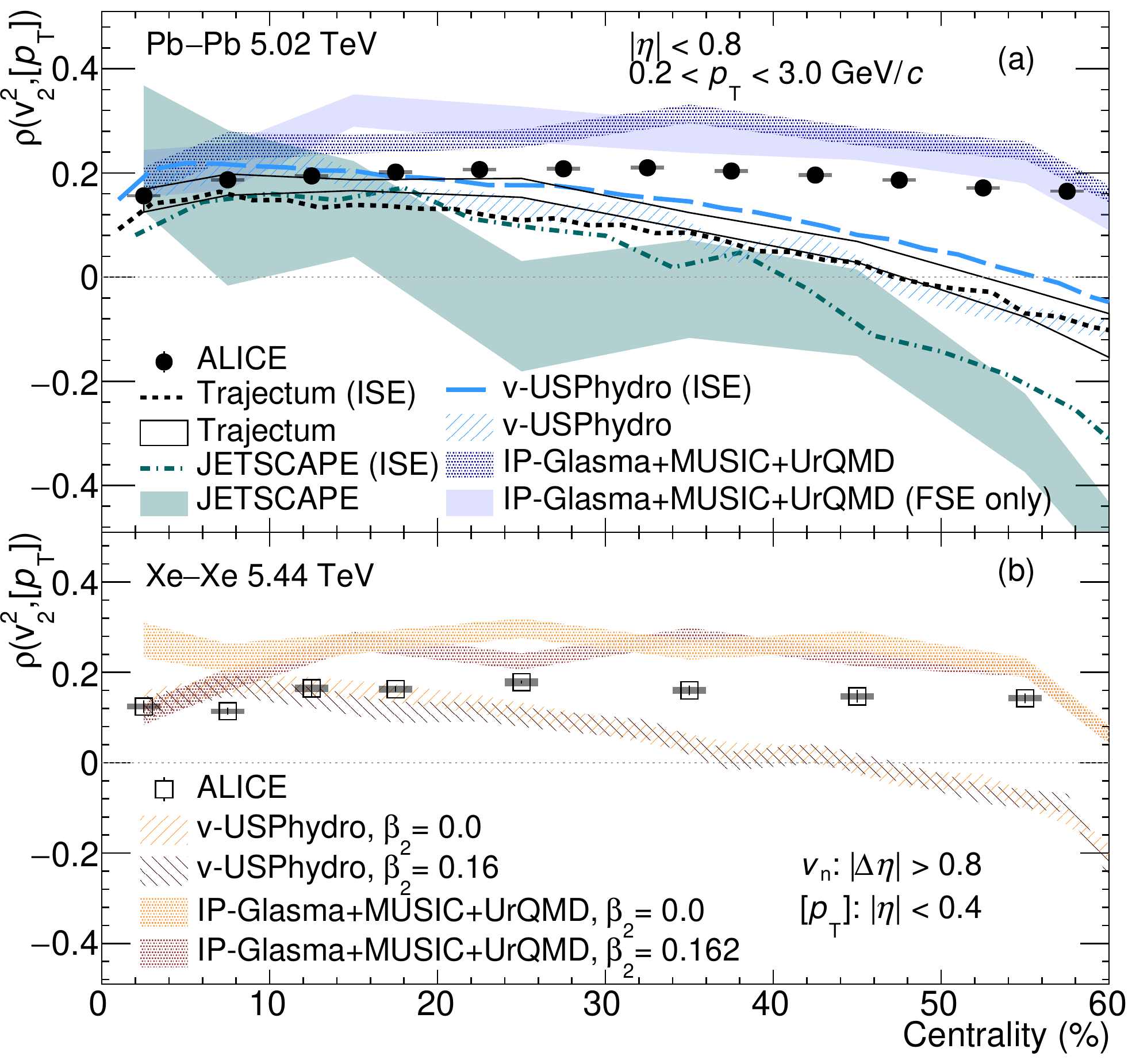}}
\caption{Correlation between the elliptic flow (azimuthal anisotropy
  of particle emission) and the mean transverse momentum of charged
  particles in Pb-Pb and Xe-Xe collisions~\cite{ref:355}.  Models
  using the Color Glas Condensate approach describe the measurement
  better than those based on \trento.}
\label{fig:correl}
\end{figure}

Initial conditions, namely the collision centrality and the transverse
distributions of nucleons within the colliding nuclei, define the
transverse energy density profile of the collision system. The average
transverse momentum of emerging particles and the Fourier coefficients
of their azimuthal distribution fluctuate event by event.  Since both
depend on the initial conditions, it is interesting to look for a
correlation between them. It turns out~\cite{ref:355} that the second
Fourier coefficient $v_2$ of the azimuthal distribution of particles
is positively correlated with the average \pt for all centralities
between zero and 60\% (Fig.~\ref{fig:correl}).
Following conclusions can be reached by comparing the measurement with
several hydro models: i) the correlation coefficient is indeed
sensitive to the initial conditions, ii) it is only weakly affected by
the subsequent stages, and iii) its magnitude and centrality
dependence are described by models based on IP-Glasma better that
those based on \trento. The second statement is supported by the
proximity of analytic initial-state estimations (ISE, lines) to the
respective full hydrodynamic calculations (shaded areas).  IP-Glasma
is the Color Glass Condensate effective theory in which the initial
conditions are dominated by the distribution of gluons. This is
different from \trento based models, in which the dominant degrees of
freedom are those of nucleons.  Understanding the initial conditions
is a prerequisite for the study of quark-gluon plasma.

\section{Hard processes: beauty production}
\label{sect:beauty}

\begin{figure}[h]
\begin{minipage}{0.44\textwidth}
\hspace*{2mm}\includegraphics[height=3.3cm]{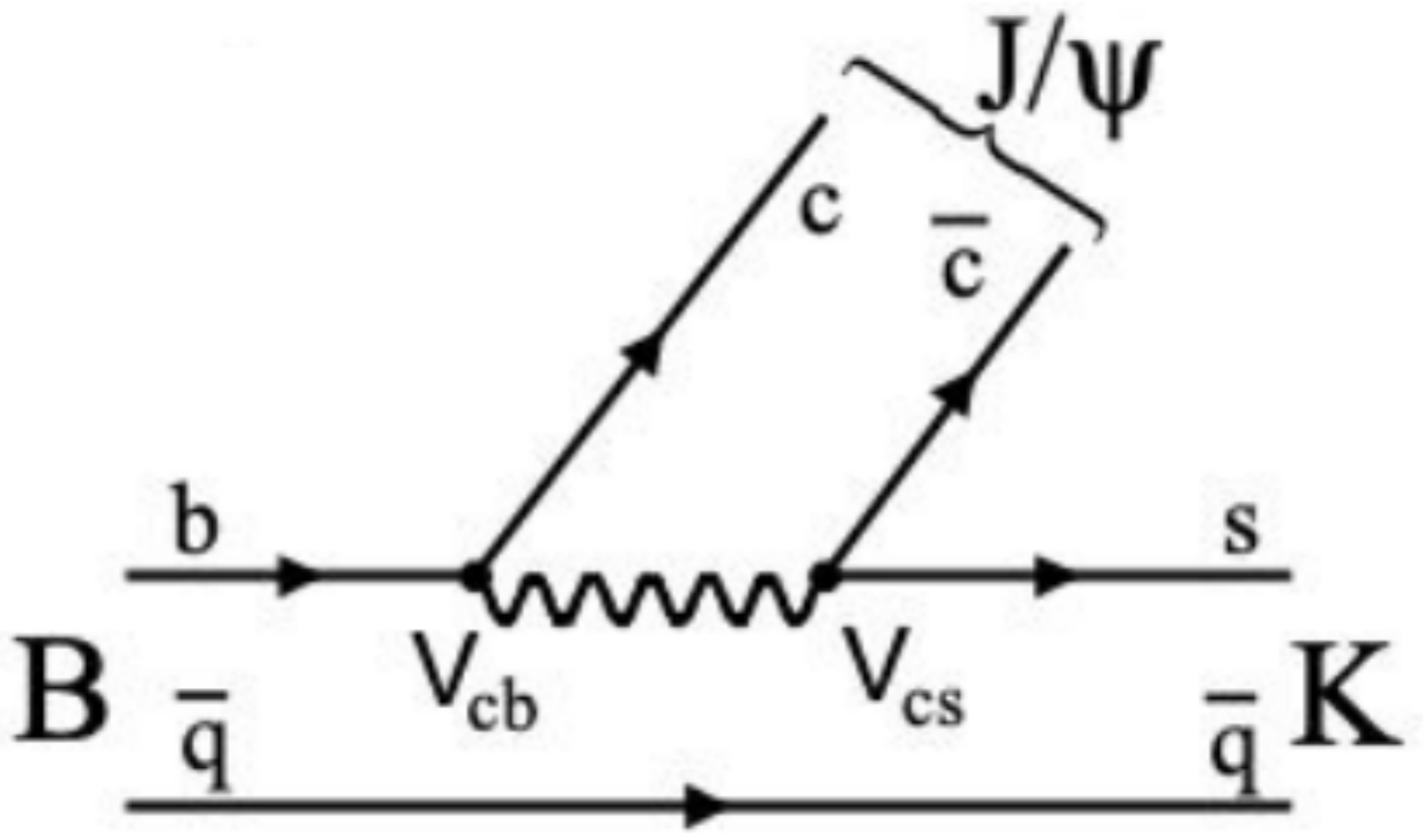}
\end{minipage}
\begin{minipage}{0.6\textwidth}
\includegraphics[height=7.8cm]{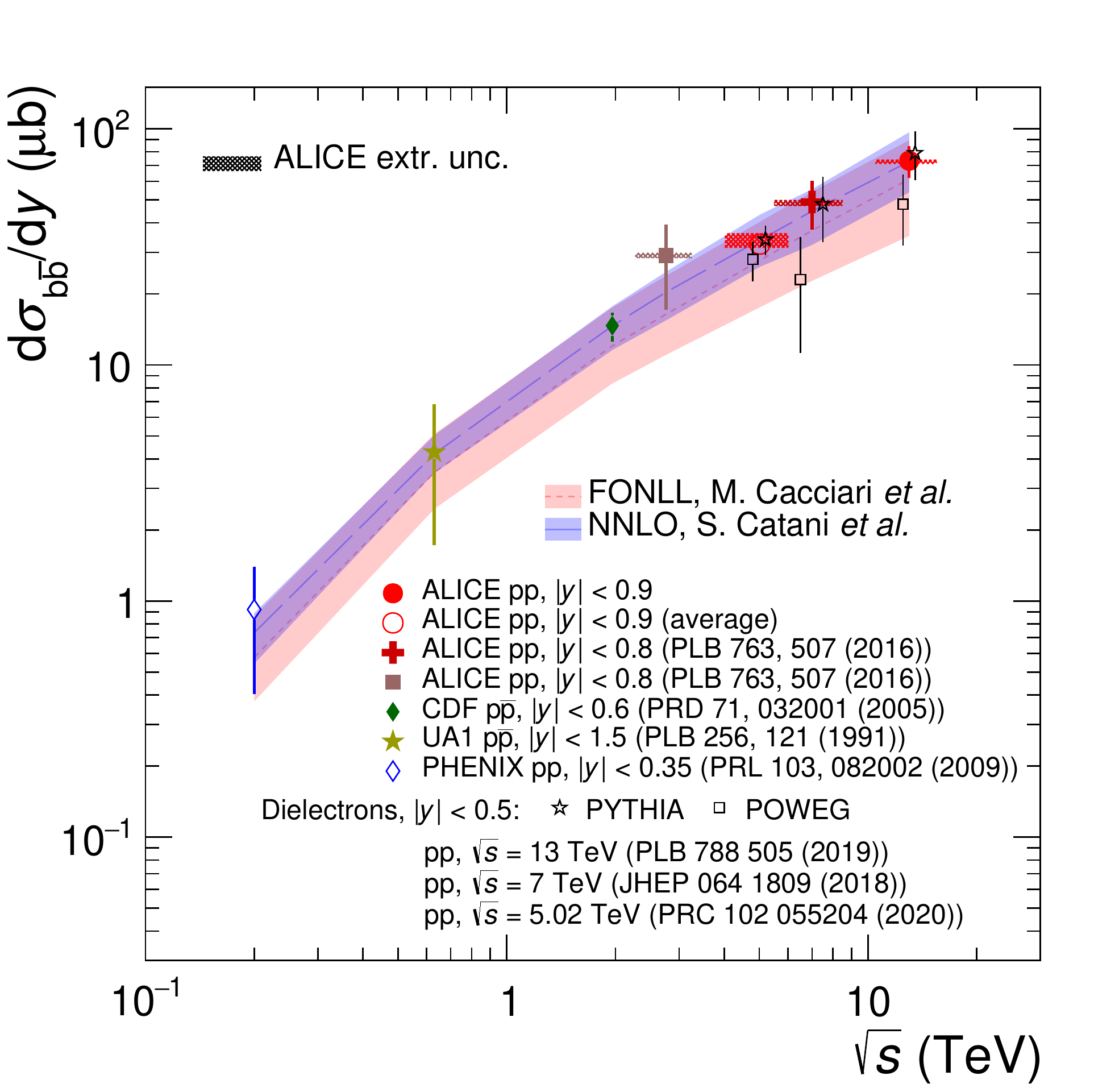}
\end{minipage}
\caption{Left: Beauty decay into non-prompt \jpsi (figure taken from
  Ref.\cite{ref:norbert}).  Right: Collision energy dependence of
  the beauty production cross section at midrapidity~\cite{ref:346}.}
\label{fig:beauty}
\end{figure}

Heavy-flavor quarks c and b decay weakly into lighter flavors. As seen
by the values of the CKM matrix elements, their most favored decays
are b$\rightarrow$c and c$\rightarrow$s. A special case of b decay
with two flavor changing vertices is its decay into \jpsi
(Fig.~\ref{fig:beauty} left).  Weak decays of s, c, and b quarks
result in characteristic hadron lifetimes of $c\tau\sim$ 2-400~cm,
0.1-0.3~mm, and 0.4-0.5 mm, respectively.  A measurement of non-prompt
(i.e. those not originating from the primary collision vertex) \jpsi
can thus be used to determine the yield of beauty hadrons. The cross
section of midrapidity beauty production in pp collisions at several
LHC energies, deduced from recent ALICE measurement of non-prompt
\jpsi, is shown in Fig.~\ref{fig:beauty} right~\cite{ref:346}. The
ALICE points match the overall collision energy systematics and
improve its precision. The \jpsi mesons were measured in the 
\jpsi$\rightarrow$e$^+$e$^-$ decay channel.  The good vertexing performance
of ALICE central barrel allowed for an efficient selection of
non-prompt \jpsi. The relation between the non-prompt \jpsi yield and
the b production cross section was taken from state-of-the-art models.
This analysis complements the ALICE measurements of beauty via non-prompt D and $\Lambda_c$. 

\section{Energy loss of c quarks}
\label{sect:closs}
\begin{figure}[h]
\center{\includegraphics[width=0.97\textwidth]{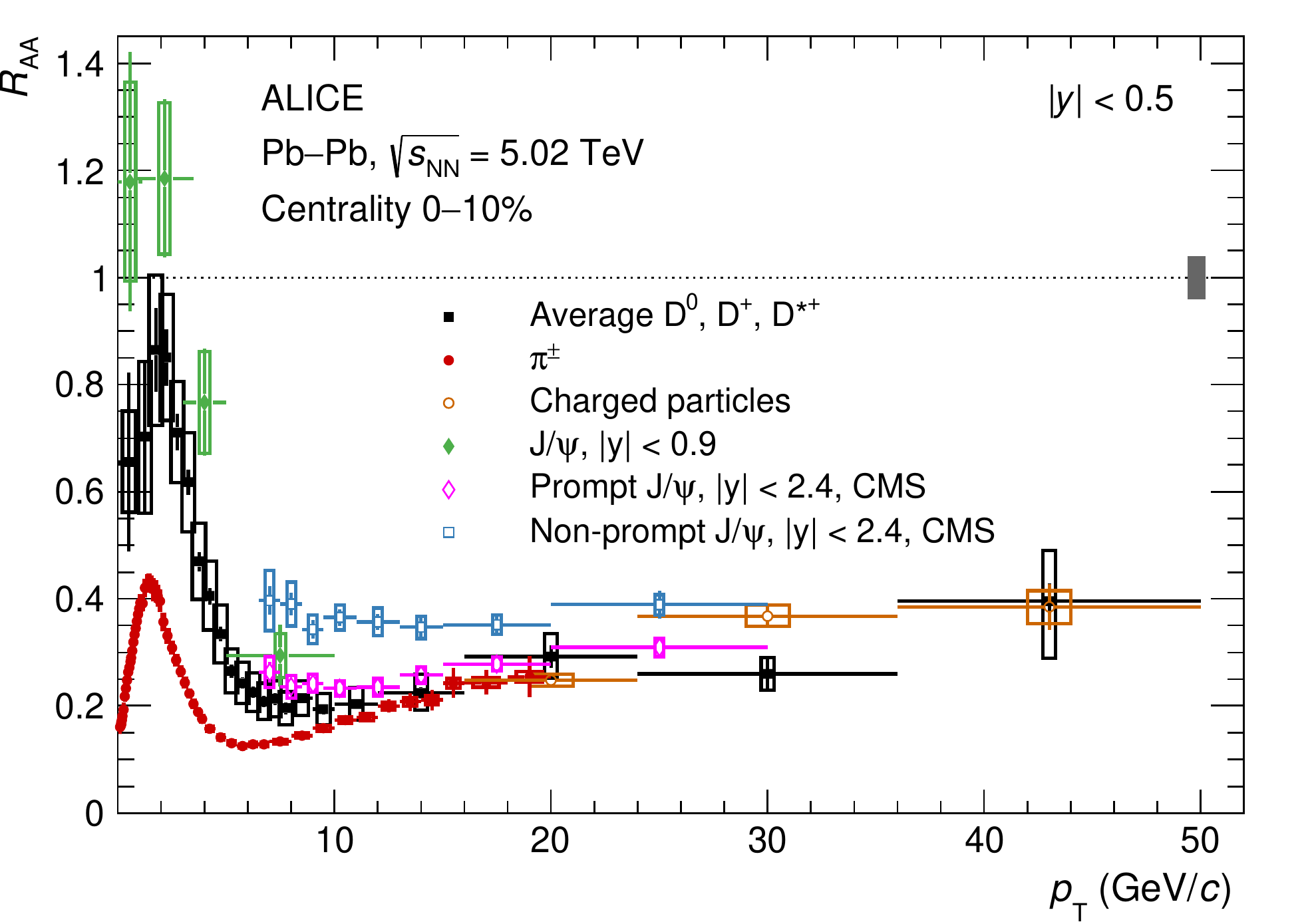}}
\caption{Nuclear modification factor of the D$^0$ yield (black points)
  compared to several other hadron species. The D$^0$ measurement
  reaches down to \pt= 0~\cite{ref:351}.}
\label{fig:raad0}
\end{figure}

Hard parton scatterings in the early stage of the Pb--Pb collision are
the main source of quarks with high mass and/or transverse momentum.
Those quarks that are moving transversally outwards faster than
average will lose part of their energy by interacting with the --
slowly expanding -- soft partonic bulk. Assuming that the emerging
hadrons have a similar velocity as their constituent quarks before
hadronization, the experimental signature should be a reduction of the
high-\pt yield in central Pb--Pb collisions compared to pp. The
characteristic \pt above which this suppression occurs should depend
on the quark mass. Figure \ref{fig:raad0} shows the nuclear
modification factor \raa, defined as the ratio of the
transverse-momentum spectrum measured in Pb--Pb collisions of a given
centrality, divided by the mean number of nucleon-nucleon collisions
$N_{\rm coll}$, to the one measured in pp.  Values below unity
indicate energy loss. The ALICE measurement of D$^0$ (black points in
Fig.~\ref{fig:raad0}) demonstrates that charm quarks interact with
medium and lose energy~\cite{ref:351}. The detailed shape of the \pt
dependence systematically changes with the hadron (or its heaviest
quark) mass.

Light and heavy hadron species get united in the limit of high \pt,
except for the non-prompt \jpsi\xspace -- a proxy for beauty -- which
for \pt = 10--20 GeV/$c$ seems to lie somewhat higher than the
rest. This suggests that b quarks at these transverse momenta lose
less energy than c quarks~\cite{ref:351,ref:363}. The quark energy
loss in this region is dominated by the radiation of gluons. A
probable reason for a reduction of the radiative energy loss of the b
quark is discussed in Section~\ref{sect:bloss}.

\begin{figure}[t]
\center{\includegraphics[width=0.99\textwidth]{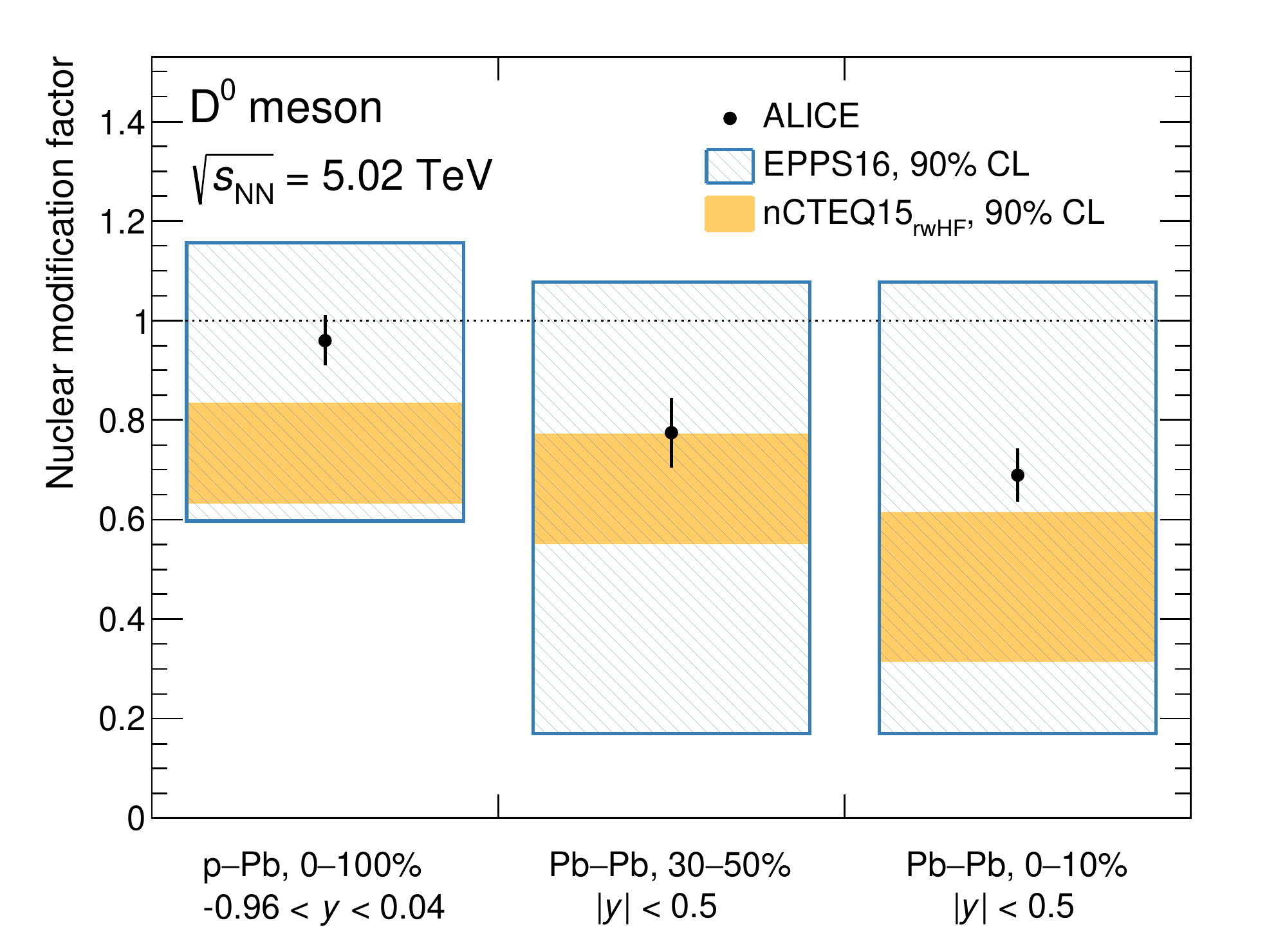}}
\caption{The \pt-integrated nuclear modification factor of the D$^0$
  yield in Pb--Pb collisions.  The total number of D$^0$ is not as
  strongly suppressed as the high-\pt part of its spectrum.}
\label{fig:intraad0}
\end{figure}

Strictly speaking, energy loss should lead to a shift to lower \pt
rather than disappearance. Since the D$^0$ has been measured down to
\pt=~0, one can integrate the spectrum and compare the total yields in
Pb--Pb and pp collisions. This is shown in
Fig.~\ref{fig:intraad0}. The \pt-integrated suppression of D$^0$ in
Pb--Pb is two times weaker than the suppression of the high-\pt part
of their spectrum. This overall suppression may come from the initial
stage: a reduced production of c quarks resulting from modifications
of the parton distribution function in Pb (shadowing).  Alternatively
it can have to do with the late stage, namely with enhanced
hadronization of the c quark into hadrons other than D$^0$. This will
be discussed in Section~\ref{sect:hadronization}.

\section{Energy loss of b quarks: dead cone}
\label{sect:bloss}
\begin{figure}[t]
\center{\includegraphics[width=1.0\textwidth]{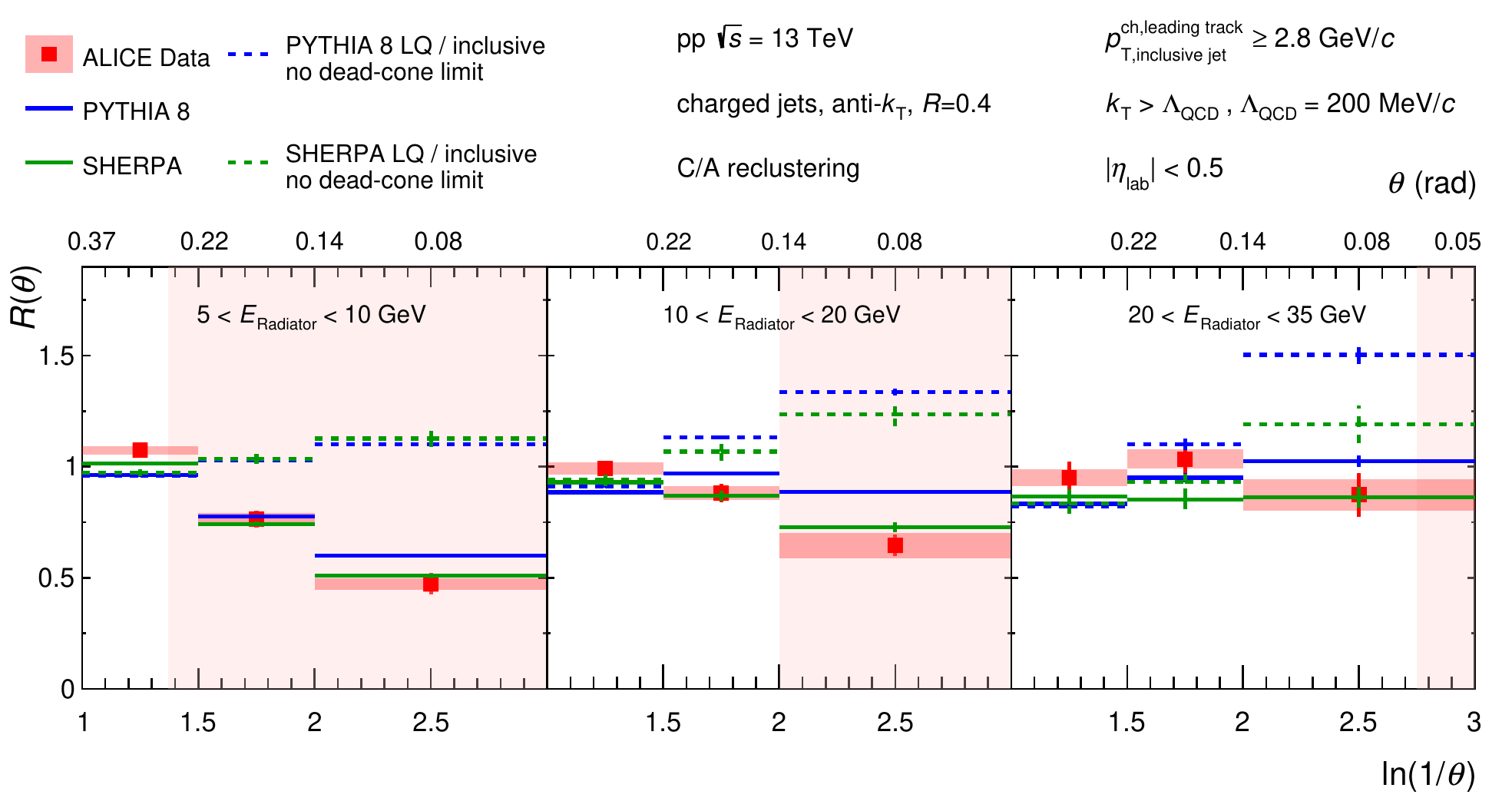}}
\caption{
The dead cone effect in the radiative energy loss of c quark in pp 
collisions, visible as a reduction of small-angle splittings~\cite{ref:335}. }
\label{fig:deadcone}
\end{figure}

Radiation of gluons by quarks, no matter whether in a colored medium
or in vacuum, is restricted to polar angles larger than $m/E$ of the
radiating quark (dead-cone effect). A detailed investigation of this
phenomenon has been performed by ALICE in pp collisions using jets
containing a D$^0$, and resulted in the first direct experimental
observation of the dead cone~\cite{ref:335}.  The D$^0$ mesons were
reconstructed in the D$^0\!\rightarrow{\rm K}^-\pi^+$ channel.  Assuming
that the c quark of D$^0$ was the leading quark of the jet, one can
reconstruct the complete history of the splittings which lead to the
jet. For this, one assumes that the c quark energy (obviously) and the
splitting angle (because of the angular ordering) decrease during the
fragmentation.  The distribution of the splitting angles in D$^0$
jets, divided by the analogous distribution in light-flavor jets, is
shown in Fig.~\ref{fig:deadcone}. The number of small-angle splittings
in jets initiated by c quarks with $5<E_{\rm Radiator}<10$~GeV (third
bin in the leftmost panel) is two times lower than in light-flavor 
jets.  The suppression of small-angle splittings gradually disappears
when going to higher quark energies (the two right panels).

\section{Hadronization in medium}
\label{sect:hadronization}
\begin{figure}[h]
\center{\includegraphics[width=0.6\textwidth]{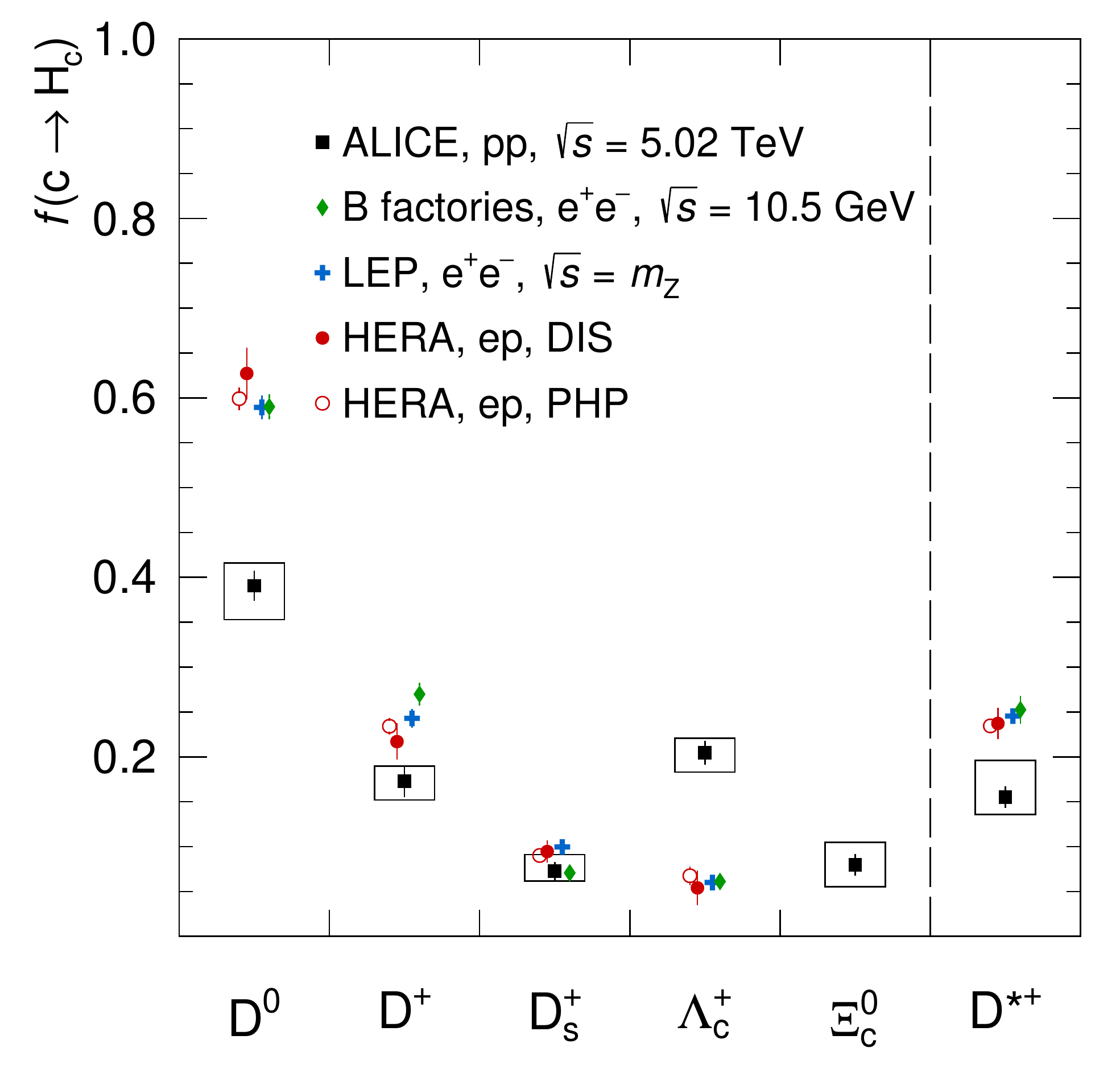}}
\caption{Charm hadronization in high energy ee, ep, and pp
  collisions~\cite{ref:334}.  The charmed baryons are significantly
  enhanced in pp as compared to ee and ep collision systems. }
\label{fig:ff5}
\end{figure}

Heavy quarks (c, b) are created in hard parton scatterings at the
beginning of the nucleus--nucleus collision and their total number is
conserved in the subsequent stages. Subject to scattering in medium,
they suffer energy loss and partially adapt their velocities to the
surrounding bulk, thus -- to some degree -- participating in its
collective motion. At the end of the partonic phase they hadronize and
emerge from the reaction as heavy mesons and baryons. Their detailed
distribution over various hadron species (fragmentation fractions) is
important to know because not always one can afford measuring all of
them to get a complete picture.  The fragmentation fractions are not
universal, as one can see already when comparing the charmed particles
emerging from ee and pp collisions. This is shown in
Fig.~\ref{fig:ff5}, in which the ALICE points represent the first
measurement of charm fragmentation fractions at the LHC and the first
measurement ever of the c fragmentation fraction to
$\Xi_c$~\cite{ref:334}. Compared to ee and ep, in pp collisions a
clear increase of the $\Lambda^+_c$ fraction and decrease of D$^0$ are
seen. This means that hadronization depends on the
environment. Models including the hadronization via coalescence with
surrounding quarks are able to describe the data. A further gradual
increase of $\Lambda^+_c$/D$^0$ is visible when going from pp to
semicentral and to central Pb--Pb collisions~\cite{ref:359}.

\newpage

\section{Closing remarks}
\label{sect:closing}
ALICE used its strengths -- low-\pt tracking, particle identification,
vertexing -- not only to study in detail various stages of Pb--Pb
collisions but also to understand the elementary QCD phenomena without
which the interpretation of the Pb--Pb data would not be possible.
This activity will be continued in the just starting (in June 2022)
Run~3 of the LHC, for which significant upgrades of the ALICE detectors
and readout were performed in the Long Shutdown 2.

\acknowledgments 
The author thanks the organizers for inviting ALICE to present its
results.  I acknowledge the creative and inspiring atmosphere and the
nice settings of the conference.

\end{document}